\documentclass[10pt,aps,pra,twocolumn,showpacs,superscriptaddress,floatfix]{revtex4-1}

\usepackage{amsmath}
\usepackage{graphicx}
\usepackage{epsfig}
\usepackage[utf8]{inputenc}
\usepackage{xfrac}
\usepackage{hyperref}
\hypersetup{colorlinks=true}

\usepackage{bbold}
\usepackage{bm}
\newcommand{\be}{\begin{equation}}
\newcommand{\ee}{\end{equation}}

\newcommand{\beq}{\begin{eqnarray}}
\newcommand{\eeq}{\end{eqnarray}}

\def\H1{\widehat{H}_1}

\newcommand{\overbar}[1]{\mkern 1.5mu\overline{\mkern-3mu#1\mkern-0mu}\mkern 1.5mu}

\usepackage{xcolor}

\begin{document}

\title[]{Non-adiabatic effects in periodically driven-dissipative open quantum systems}

\author{Viktor Reimer}
\affiliation{Institute for Theory of Statistical Physics, RWTH Aachen University, 52056 Aachen, Germany}

\author{Kim G. L. Pedersen}
\affiliation{Institute for Theory of Statistical Physics, RWTH Aachen University, 52056 Aachen, Germany}

\author{Niklas Tanger}
\affiliation{Institute for Theory of Statistical Physics, RWTH Aachen University, 52056 Aachen, Germany}

\author{Mikhail Pletyukhov}
\affiliation{Institute for Theory of Statistical Physics, RWTH Aachen University, 52056 Aachen, Germany}

\author{Vladimir Gritsev}
\affiliation{Institute for Theoretical Physics, Universiteit van Amsterdam, 1098 XH Amsterdam, The Netherlands}
\affiliation{Russian Quantum Center, 143025 Skolkovo (Moscow), Russia}


\begin{abstract}
We present a general method to calculate the quasi-stationary state of a driven-dissipative system coupled to a transmission line (and more generally, to a reservoir) under periodic modulation of its parameters. Using Floquet's theorem, we formulate the differential equation for the system's density operator which has to be solved for a single period of modulation. On this basis we also provide systematic expansions in both the adiabatic and high-frequency regime.
Applying our  method to three different systems -- two- and three-level models as well as the driven nonlinear cavity -- we propose periodic modulation protocols of parameters leading to a temporary suppression of effective dissipation rates, and study the arising non-adiabatic features in the response of these systems.

\end{abstract}

\maketitle


\section{Introduction}

Classical Floquet theory~\cite{F} gave long-standing inspiration for studies of a variety of time-periodic processes in nature and has found a huge domain of applicability in fields ranging from dynamical system's theory to technology. In quantum mechanics, Bloch's theorem for crystals~\cite{Bloch} represents the momentum-space analogue of Floquet's seminal work, while in time-domain the concept of quasi-energy was introduced only in the 1960's by Zeldovich~\cite{Zeldovich}.

Periodic time-dependent processes are natural in quantum optics where the input laser field provides a fast periodic driving of the system. To the best of our knowledge, it was Shirley~\cite{Shirley} who first applied a Floquet formalism in quantum optics. He clarified the connection between a semi-classical external field drive and its quantized strong, resonant single-mode field counterpart applied to an $N$-level atom based on general considerations using Floquet's theorem. Various extensions of this work, which focused on the semi-classical picture suggested by the strong-intensity nature of laser fields, have been reviewed in~\cite{Chu1, MFSF, Chu2}.

In the quantum regime, studies of periodically driven-dissipative (open) quantum systems -- immediately relevant for quantum optics -- have lead to a whole new class of physics inaccessible in equilibrium systems. Most of the earlier developments are reviewed in~\cite{GH, Zeldovich2}, including the paradigmatic two-level systems, tunneling problems and spin-boson models. More recent examples with potential for technological applications cover the emergence of topological phases -- so-called Floquet topological insulators~\cite{Cayssol13} -- non-thermal steady states exhibiting localization~\cite{Khemani16} and artificial gauge systems~\cite{Aidelsburger17}.

A particularly useful approach for studying driven-dissipative systems is the so-called Floquet-Liouvillie approach~\cite{Ho86} which reduces to a Lindbladian master equation under the Floquet-Markov approximation~\cite{floquet-markov}. We note that there is a subtlety concerning different procedures of performing the Markovian approximation. In general, the Markovian approximation for the eigenenergy spectrum performed on the level of the undriven Hamiltonian differs from performing it on the level of the driven Floquet quasi-energy spectrum. Applications of such approaches, which have shown to capture some interesting features of periodically driven-dissipative quantum systems, range from transport problems such as electronic pumping~\cite{Janine12} to dynamical decoupling schemes for qubit control~\cite{Fn6}, see also~\cite{KLH} for a review. However, due to the large separation of system and driving time-scales -- a regime where the Markovian approximation is very well valid -- it has been most widely applied in the context of quantum optics, see e.g.~\cite{davies-spohn, BP, KDH, Fn1, Fn5, DSY, per-thermo, Kosloff, modulated}.

Recently, there has been increasing interest in investigating driven-dissipative phase transitions under time-periodic driving, e.g. for the Rabi model~\cite{Boite17}. We will later focus on a different system exhibiting a dissipative phase transition, the Kerr nonlinearity model, which has been analytically solved for the stationary case by Drummond and Walls~\cite{Drummond80} in the 1980's. It has been shown experimentally, that the bistable behavior of Kerr nonlinearities can be exploited to confine the manifold of available states in superconducting qubits to coherent states under special two-photon driving schemes~\cite{Leghtas15}. Since then, driving of this model has been subject to extensive theoretical studies~\cite{Ciuti17, Casteels16, BartoloCiuti, Rodriguez17}.  

Analytical investigations employing the Floquet-Liouville approach are in practice restricted to either adiabatic or high-frequency limits, and only for few problems~\cite{Dion, YE} it is feasible to derive closed systems of equations. Whereas for closed systems the high-frequency Magnus expansion is standardly -- and successfully -- used~\cite{Polkovnikov15}, the complex eigenenergies characteristic for open systems prohibit any truncation of the Magnus series as it typically yields exponentially increasing, i.e. unphysical, terms. On the other hand, an adiabatic approximation may be invalid even for slow driving frequencies if the effective dissipation rate is (temporarily) suppressed. This will be the case for models discussed in this paper and we will show that {\it non-adiabatic effects} become prominent even when the modulation is slow as compared to bare dissipation rates.  

In this paper, we establish a general framework for studying the quasi-stationary regime of periodically driven-dissipative quantum systems that is capable of systematically addressing both slow and fast modulations. It extends the previously developed method based on the scattering formalism~\cite{Pletyukhov17}, allowing us to capture multi-photon processes via the equation of motion approach. The latter
is designed in such a way that an integration is required over a single period of modulation only. The adiabatic and high-frequency limits can therefore be efficiently benchmarked against exact numerical results. We apply our framework to investigate
non-adiabatic effects which in general arise due to a nearly vanishing Liouvillian gap. These effects can appear useful for implementing adiabatic quantum computation with superconducting qubits coupled to baths~\cite{Barends}, and for various dynamical decoupling schemes~\cite{decoupling}. 

We apply our approach to three quantum optical systems exhibiting a critical suppression of the smallest dissipation rate. In section~\ref{sec:mollow}, a two-level system with a periodically driven coupling to the transmission line is considered. This model exhibits the striking feature of alternating in time between bunching and anti-bunching statistical behavior of reflected photons.

In section~\ref{sec:eit} we show that similar non-adiabatic effects can be realized with a three-level $\Lambda$-system when the drive field intensity is periodically modulated.

Finally, in section~\ref{sec:kerr} we consider the Kerr nonlinearity model where we focus on the system's response to changing parameters across the region of the dissipative phase transition and the emergence of the hysteretic behavior which has been recently theoretically predicted~\cite{BartoloCiuti} and experimentally observed~\cite{Rodriguez17}.

\section{Model}\label{sec:model}

\begin{figure}
	\centering
	\includegraphics[width=\columnwidth]{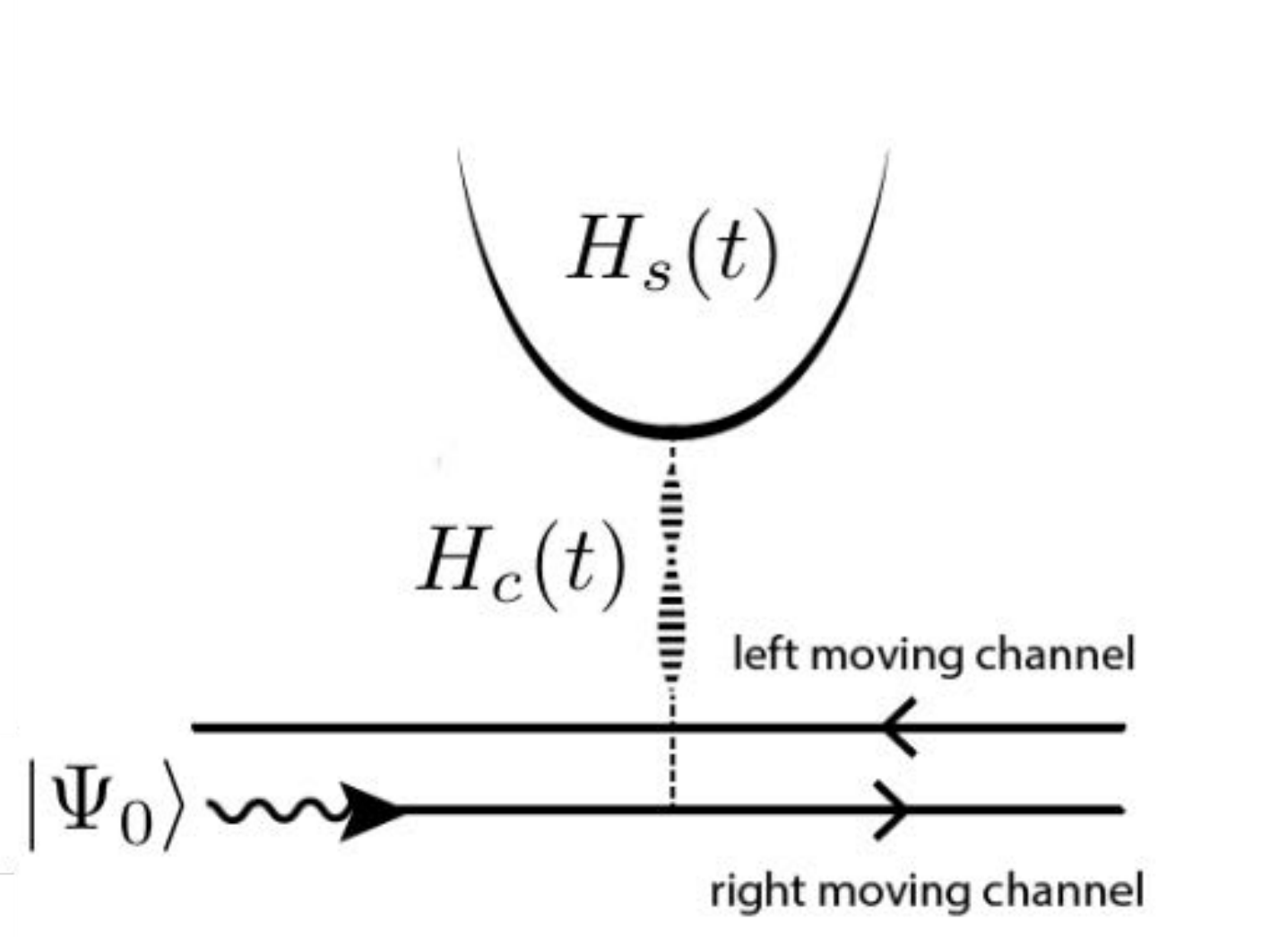}
	\caption{Open quantum system model: A local system is coupled to a transmission line supporting left- and right-propagating modes. Either the system's parameters  or its coupling to the transmission line is periodically modulated. The input pulse into the transmission line is given by a coherent state $| \Psi_0 \rangle$ in the right-propagating mode $\omega_0$. The pulse's intensity is characterized by the photonic flux $f$.}
	\label{fig:coupling_setup}
\end{figure}

The models considered within this paper all share the notion of a quantum system described by the local Hamiltonian $H_{s}(t)$ which is driven via a coupled transmission line, or waveguide, by a coherent pulse $| \Psi_0 \rangle$ characterized in terms of the photonic flux $f$, as shown in Fig.~\ref{fig:coupling_setup}. The whole setup is described by the Hamiltonian
\begin{align}
	H(t) = H_{s}(t) + H_{w} + H_{c}(t), \label{eq:general_hamiltonian}
\end{align}
with the waveguide contribution $H_{w} = \sum_{\alpha} \int d\omega (\omega_0 + \omega) a^\dagger_{\alpha \omega}a_{\alpha \omega}$ written in terms of left- and a right-propagating fields labeled by mode ($\omega$) and direction ($\alpha=L,R$) indices. We either assume a time-dependent coupling strength $g(t)$ in the coupling Hamiltonian
\begin{align}
H_{c} (t) &= \sum_\alpha \int d\omega \left[  \frac{g(t)}{\sqrt{2}} a^\dagger_{\alpha \omega} O + h.c. \right], \label{eq:general_coupling}
\end{align}
where $O$ is some operator of the local system, or a periodic modulation of some parameters of the local quantum system Hamiltonian $H_{s}(t)$ itself.

The form of $H_{w}$ implies the general assumption that the dispersion of the transmission line can be linearized around a working frequency $\omega_0$, such that $\omega_k = \nu (k - k_0) + \omega_0$, where $\nu$ is the group velocity. For convenience, we employ units in which $\nu = \hbar = 1$ holds. Extending the linearized dispersion to the full spectrum is known as the wide band approximation and is valid if the working frequency $\omega_0$ is large compared to all other energy scales, including the driving frequency, $\omega_0 \gg \Omega$. Note that this also falls in line with the rotating wave approximation (RWA) leading to the coupling Hamiltonian~\eqref{eq:general_coupling} and effectively constitutes the Markovian limit which holds even in the case of time-periodic modulation.

Time dynamics of the  system's reduced density matrix is then governed by the Lindblad master equation
\begin{align}
\dot{\rho}(t) &= -i \left[H_{\textrm{eff}}(t), \rho(t)\right] + \gamma(t) \mathcal{D}\left[O\right] \rho(t), \label{eq:lind}
\end{align}
with a time-dependent dissipation rate $\gamma(t) \equiv \pi |g(t)|^2$, and
\begin{align}
H_{\textrm{eff}}(t) &= H_s(t) + \sqrt{\pi f} g(t) O + \sqrt{\pi f} g^\ast(t) O^\dagger, \\
\mathcal{D}[O] \rho(t) &= O \rho(t) O^\dagger - \frac{1}{2} O^\dagger O \rho(t) - \frac{1}{2} \rho(t) O^\dagger O.
\end{align}

\section{Quasi-stationary approach}
\label{sect:quasi_st}

The aim of this section is to set up a formalism, which allows us to directly access the quasi-stationary solution of~\eqref{eq:lind} in the long-time limit, using Floquet's theorem in the time representation. Traditionally (see, e.g.~\cite{Boite17}, for a recent application), Floquet's theorem is employed to get rid of an explicit time-dependence of periodic Hamiltonians or Liouvillians by switching to the Fourier representation. The problem is thereby reduced to a static eigenvalue problem for the so-called Floquet quasi-energies and modes in an enlarged Hilbert space. While this procedure is in principle always possible, it introduces certain difficulties for practical numerical calculations, since it necessitates a truncation of the infinite number of Floquet modes. This is especially perilous if an additional cut-off, e.g. in the Fock basis, is required as for example in the case of the Kerr nonlinearity model considered in section~\ref{sec:kerr}. For this reason, we prefer a formulation in terms of differential equations for quasi-stationary states which can be solved on a single period of modulation.  

The starting point of our consideration is the master equation for the reduced density operator $\rho(t)$ of the local system,
\begin{align}
	\frac{d}{d t} \rho (t) = - i L (t) \rho (t), \label{eq:vonNeumann}
\end{align}
with the Liouvillian superoperator $L(t)$ generalizing the one in \eqref{eq:lind}. In spite of the time-dependence, it must have a zero eigenvalue, as it is dictated by the trace preservation of $\rho (t)$. It is convenient to explicitly split off the corresponding zero-eigenmode of $L(t)$. In order to do so, we fix some matrix representation of $\rho(t)$ and express the occupation probability of the ground state by  $\rho_{00} (t)= 1- \sum_{i=1}^{N-1} \rho_{ii} (t)$, where $N$ is the number of states in the system. In the vectorized form, i.e. by re-stacking the columns of the matrix representation of $\rho(t)$ into an $N^2$-dimensional vector $(\rho_{00}, \vec{\rho})^{T}$, the master equation~\eqref{eq:vonNeumann} turns into
\begin{align}
	\frac{d}{d t} \left( \begin{array}{c} \rho_{00} (t) \\ \vec{\rho} (t) \end{array} \right)
	= 
	\left( 
		\begin{array}{cc} 
	        -i L_{00} (t) & \vec{\bar{C}}^T (t) \\ 
		    \vec{C} (t) & -i \bar{L} (t)  
		\end{array} 
	\right)  
	\left( 
		\begin{array}{c} 
		    \rho_{00} (t) \\ 
		    \vec{\rho} (t) 
		\end{array} 
	\right).
\end{align}
As a result, all information has been encoded in the $(N^2-1)$-dimensional state vector $\vec{\rho} (t)$ governed by the differential equation
\begin{align}
	\frac{d}{d t} \vec{\rho} (t) 
		&= 
		[-\vec{C} (t) \otimes \vec{E}^T - i  \bar{L} (t)] \vec{\rho} (t) + \vec{C} (t) \nonumber \\
		& \equiv 
		A (t) \vec{\rho} (t) + \vec{C} (t), \label{eq:dot_rhop}
\end{align}
where $\vec{E}$ consists of $1$'s ($0$'s) in the positions corresponding to the diagonal (off-diagonal) elements of $\vec{\rho}$. 

For  a time-independent Liouvillian, Eq.~\eqref{eq:dot_rhop} allows for a direct calculation of the true steady state $\vec{\rho}_{\textrm{st}} = - A^{-1} \vec{C}$. In the case of a time-periodic driving, time translational invariance is lost even in the long-time limit and the time-dependent quasi-stationary state $\vec{\rho}_{\textrm{qs}} (t)$ will essentially follow the persistent external modulation after some transient time-regime in which the influence of the initial state gradually decays. We are interested in this long-time limit and take the initial time $t_0 = - M_0 T \to - \infty$ to be in the far past, where we assume without loss of generality that it is back by a large integer multiple $M_0 \gg 1$ of the driving period $T$. The ansatz
\begin{align}
	\vec{\rho} (t) = \lim_{t_0 \to - \infty} O (t) \vec{\rho} (t_0) + \vec{\rho}_{\textrm{qs}} (t)
	\label{ansatz}
\end{align}
reflects the split structure of Eq.~\eqref{eq:dot_rhop} and gives a clear physical interpretation of the appearing vectors and matrices.

The matrix $O(t)$ describes the gradual decay of the initial conditions in the far past and is solely governed by the periodic matrix $A(t)$,
\begin{align}
  \dot{O} (t) &= A (t) O (t), 
  	& O (0) &= \mathbb{1}.
\label{Ot}
\end{align}
Note that the reference time has been shifted from $t_0$ to zero which is possible here due to the periodic nature of $A(t)$. According to Floquet's theorem, the solution of this differential equation can be represented as 
\begin{align}
O (t) = P (t) e^{B t}, \label{Ot0}
\end{align}
where $P(t)=P(t+T)$ is a periodic matrix function with the initial condition $P(0) = \mathbb{1}$. The constant matrix $B$, which is obtained from the monodromy matrix $O (T) =  P (T) e^{BT} = e^{BT}$, has eigenvalues with negative real parts such that $\lim_{t \to \infty} O(t) = 0$ holds, and all information about initial conditions in Eq.~\eqref{ansatz} is lost as required.

After the initial conditions have fully decayed, only the time-periodic quasi-stationary state vector $\vec{\rho}_{\textrm{qs}} (t)$ remains. It is governed by the differential equation
\begin{align}
  \dot{\vec{\rho}}_{\textrm{qs}} (t) &= A (t) \vec{\rho}_{\textrm{qs}} (t) + \vec{C} (t), 
  	& \lim_{t_0 \to - \infty} \vec{\rho}_{\textrm{qs}} (t_0) &=0,
	\label{ct}
\end{align}
where unlike in the case of $O(t)$ the reference time $t_0$ remains unaltered to account for the fact that we are interested in the long-time limit. The differential equation~\eqref{ct} can be formally integrated to
\begin{align}
 \vec{\rho}_{\textrm{qs}} (t) &= O (t) \int_{- \infty}^t d t' O^{-1} (t') \vec{C} (t').
\label{ct0}
\end{align}
Periodicity of this solution is straightforwardly seen from equation~\eqref{Ot0} and the periodicity of $P(t)$ and $\vec{C}(t)$, and therefore it is sufficient to study its behavior on the finite interval $\tau_c \in [0,T]$.

To further evaluate~\eqref{ct0}, we first split the integration range into two intervals $[-\infty,0]$ and $[0, \tau_c]$,
\begin{align}
	\vec{\rho}_{\textrm{qs}} (\tau_c) &= O (\tau_c) \int_{-\infty}^{0} d t' O^{-1} (t') \vec{C} (t') +  \vec{c} (\tau_c), 
\label{sqs0}
\end{align}
where
\begin{align}
	\vec{c}(\tau_c) = O (\tau_c) \int_{0}^{\tau_c} d t' O^{-1} (t') \vec{C} (t') \label{eq:cvec}
\end{align}
is defined in analogy with Eq.~\eqref{ct0} with the reference time shifted to zero. We note that instead of inverting the large matrix $O(t)$ appearing in Eq.~\eqref{eq:cvec}, it is more favourable to instead numerically solve the differential equation
\begin{align}
  \dot{\vec{c}} (t) &= A (t) \vec{c}(t) + \vec{C} (t),
  	& \vec{c} (0) &=0.
	\label{ct_ode}
\end{align}

Next, the interval $[-\infty , 0]$ is split into an infinite number of intervals $[-(n+1) T, -n T]$, $n \in \mathbb{N}_0$. Using the periodicity of $P (t)$, we represent the first term of Eq.~\eqref{sqs0} by a geometric progression with the factor $e^{B T}$. Resumming it, we obtain 
\begin{align}
  \vec{\rho}_{\textrm{qs}} (\tau_c) 
  &= O (\tau_c) (1-O (T))^{-1} \vec{c} (T) + \vec{c} (\tau_c).
\label{sqsO}
\end{align}
Thus, to evaluate $\vec{\rho}_{\textrm{qs}} (t)$, it is sufficient to solve the set of equations~\eqref{Ot} and~\eqref{ct_ode} on the finite interval $0 \leq \tau_c \leq T$. In fact, the solution~\eqref{sqsO} obeys the differential equation~\eqref{ct} with periodic boundary conditions rather than the initial condition therein.
 
\subsection{Adiabatic expansion}
\label{sect:adiab}

In the adiabatic limit, the external driving of parameters is sufficiently slow such that the state can instantaneously adapt to its new environment, $\rho_{\textrm{qs}} (t) \approx - A^{-1} (t) \vec{C} (t) \equiv \rho_{\textrm{inst}} (t)$. 

In order to consistently compute adiabatic corrections to the instantaneous solution $\rho_{\textrm{inst}} (t)$, we insert the relation $O^{-1}(t) = - \frac{d}{dt} [ O^{-1}(t) ]  A^{-1}(t)$ into Eq.~\eqref{ct0}.  Integrating it by parts we obtain
\begin{align}
\vec{\rho}_{\textrm{qs}} (t) &=  \rho_{\textrm{inst}} (t) \nonumber \\
&\phantom{=} - O (t) \int_{-\infty}^t d t' O^{-1} (t') \frac{d}{d t'} \rho_{\textrm{inst}} (t').
\end{align}
Iterating this procedure leads to a geometric series that can  be resummed to
\begin{align}
	\vec{\rho}_{\textrm{qs}} (t)
	&=  \frac{1}{1-A^{-1}(t)\frac{d}{dt}} \vec{\rho}_{\textrm{inst}} (t) \nonumber \\
	&\approx \vec{\rho}_{\textrm{inst}} (t) + A^{-1} (t) \frac{d}{d t}  \vec{\rho}_{\textrm{inst}} (t). \label{eq:adiabatic_sol}
\end{align}
We note that the convergence of this series relies on some sort of an adiabaticity condition. If such a condition is violated or generally not provided, the adiabatic expansion~\eqref{eq:adiabatic_sol} breaks down.

\subsection{High-frequency expansion}
\label{sec:high_fr}

The Magnus expansion is frequently used for analyzing high-frequency processes in driven quantum optical systems. Note, however, that it is originally designed for applications in closed systems where the evolution is unitary. For driven-dissipative systems with Liouvillian dynamics, it often produces --  according to our experience -- exponentially growing, unphysical terms already in the first order of expansion.  

Instead of the Magnus expansion, we perform a straightforward high-frequency expansion of the master equation~\eqref{eq:dot_rhop} in the following way. Since in the quasi-stationary regime $A (t)$, $C (t)$ and $\vec{\rho} (t)$ are all periodic functions of time, let us explicitly split off the constant zero-frequency component for each of these objects
\begin{align}
X (t) &= \overbar{X} + \widetilde{X} (t), \quad  \overbar{X}  = \frac{1}{T} \int_0^T dt' X (t'). \label{eq:freq_split}
\end{align}
Here $\widetilde{X} (t)$ is a periodic function with zero time average. Then, we rewrite the master equation~\eqref{eq:dot_rhop}, which must also hold in the quasi-stationary regime with periodic boundary conditions, as (vector notation omitted in the following)
\begin{align}
	\frac{d}{dt} \widetilde{\rho} (t) 
	&=
	\left( \overbar{A} + \widetilde{A} (t)\right) \overbar{\rho} + \left( \overbar{A} + \widetilde{A} (t)\right) \widetilde{\rho} (t) + \overbar{C} + \widetilde{C} (t).
\label{eq2}
\end{align}
The constant average $\overbar{\rho}$ can be expressed in terms of the periodic part $\widetilde{\rho}$ if one integrates Eq.~\eqref{eq2} over one period,
\begin{align}
\overbar{\rho} = - \overbar{A}^{-1}  \left( \overbar{C}+ \frac{1}{T} \int_0^T d t' \widetilde{A} (t') \widetilde{\rho} (t')  \right).
\label{sol_mean}
\end{align}
Now, perform a high-frequency expansion of $\rho (t)$ in powers of the inverse modulation frequency $\Omega = 2\pi /T$ 
\begin{align}
\overbar{\rho} = \sum_{n=0}^{\infty} \frac{1}{\Omega^n} \overbar{\rho}^{(n)}, && \widetilde{\rho} (t) =  \sum_{n=1}^{\infty} \frac{1}{\Omega^n} \widetilde{\rho}^{(n)} (t).
\label{eq:high_fr}
\end{align}
The hierarchy of differential equations resulting from this ansatz,
\begin{subequations}
\label{eq:highfreqsol}
\begin{align}
	\frac{d}{dt} \widetilde{\rho}^{(1)} (t) 
	& =
	\widetilde{A} (t) \overbar{\rho}^{(0)}  + \widetilde{C} (t), \label{1osc}
	\\
	\frac{d}{dt} \widetilde{\rho}^{(n)} (t) 
	& = 
	\widetilde{A} (t) \overbar{\rho}^{(n-1)} +\overbar{A} \widetilde{\rho}^{(n-1)} (t) 
	\nonumber \\
	& \phantom{=}
	+ \widetilde{A} (t) \widetilde{\rho}^{(n-1)} (t)
	\nonumber \\
	& \phantom{=}
	- \frac{1}{T} \int_0^T d t' \widetilde{A} (t') \widetilde{\rho}^{(n-1)} (t')  , \quad n \geq 2, 
\end{align}
\end{subequations}
can be iteratively solved as follows. First, we extract from Eq.~\eqref{sol_mean} the leading order of the expansion for the constant average
\begin{align}
\overbar{\rho}^{(0)} = - \overbar{A}^{ -1} \overbar{C}, \label{eq:rhoav0}
\end{align}
with which we can formally solve Eq.~\eqref{1osc}
\begin{align}
\widetilde{\rho}^{(1)} (t) &= -  \left( \int_0^{t} d t'  \widetilde{A} (t') -\frac{1}{T} \int_0^T d t  \int_0^{t} d t'  \widetilde{A} (t') \right) \overbar{A}^{-1}   \overbar{C} \nonumber \\
 &\phantom{=} + \int_0^{t} d t'  \widetilde{C} (t') -\frac{1}{T} \int_0^T d t  \int_0^{t} d t'  \widetilde{C} (t').
\end{align}
Knowing $\widetilde{\rho}^{(1)} (t)$, we can then also extract $\overbar{\rho}^{(1)}$ from Eq.~\eqref{sol_mean}:
\begin{align}
\overbar{\rho}^{(1)} = - \overbar{A}^{-1}   \frac{1}{T} \int_0^T d t' \widetilde{A} (t') \widetilde{\rho}^{(1)} (t'). \label{eq:rhoav1}
\end{align}
The higher-order contributions are obtained by an analogous  iterative procedure.

\section{Driven two-level system}\label{sec:mollow}

Here we apply the Floquet formalism developed above to a setup in which the local quantum system has two levels (a qubit) and the coupling to the transmission line is periodically modulated. We have already discussed this setup in the recent publication~\cite{Pletyukhov17} in the regime of weak intensities $f \ll \gamma$ of the coherent input pulse using Floquet scattering theory. The present approach allows us to extend our previous results to larger input powers $f \geq \gamma$.

The Hamiltonian~\eqref{eq:general_hamiltonian} of this system is  specified by $H_s (t) = \omega_e  \sigma_+ \sigma_-$ and $O = \sigma_-$. Going to the co-rotating frame, we find that the master equation~\eqref{eq:dot_rhop} for $\vec{\rho} (t) \to \vec{s} (t) = \langle \hat{\vec{s}} (t) \rangle \equiv (\langle \tilde{\sigma}_{+} (t) \rangle, \langle \tilde{\sigma}_{-} (t) \rangle, \langle 1+ \sigma_z (t)\rangle  )^T$
uses
\begin{align}
A (t) = \left( \begin{array}{ccc} 
- i \delta  - \gamma (t)/2 & 0 & - i \sqrt{\pi f} g (t) \\
0 &  i \delta  - \gamma (t)/2 &  i \sqrt{\pi f} g^* (t) \\
-2 i \sqrt{\pi f} g^* (t) & 2 i \sqrt{\pi f} g (t) & - \gamma (t)
\end{array} \right),
\label{At}
\end{align}
and
\begin{align}
\vec{C} (t) = (i  \sqrt{\pi f} g (t),-i \sqrt{\pi f} g^* (t), 0)^T .
\label{Ct}
\end{align}
Here we introduced the detuning $\delta  = \omega_0 - \omega_{e} $ as well as  $\langle \tilde{\sigma}_{\mp}  (t) \rangle  =\langle \sigma_{\mp} (t) \rangle e^{\pm i \omega_0 (t-t_0)}$. 

Importantly, in a broad range of $f$, the smallest dissipation rate is solely determined by the coupling strength $g$, and quenching $g \to 0$ will cause a critical slowing down of the system's Liouvillian dynamics. We exploit this property to design a modulation protocol $g (t)$ aiming to achieve time-intervals where the modulation frequency $\Omega =2 \pi/T$ exceeds the scale set by the smallest dissipation rate, $\Omega > \gamma_{\textrm{min}} (t)$. Within these time-intervals, we expect the system's response to be non-adiabatic such that the expansion~\eqref{eq:adiabatic_sol} breaks down.

\subsection{Reflection and transmission}

Applying the standard input-output relations, we find reflection and transmission amplitudes 
\begin{align}
\mathcal{R}(t) \equiv \frac{\langle a_{L, \textrm{out}}(t) \rangle}{\langle a_{R, \textrm{in}}(t)\rangle} &= -i \sqrt{\frac{\pi}{f}} g(t) s_2(t), 
\label{Rt} \\
\mathcal{T}(t) \equiv \frac{\langle a_{R, \textrm{out}}(t) \rangle}{\langle a_{R, \textrm{in}}(t)\rangle} &= 1 + \mathcal{R}(t),
\label{Tt}
\end{align}
which are expressed via the second component of the vector $\vec{s}(t)$.

\begin{figure}[t]
	\centering
	\includegraphics[width=\columnwidth]{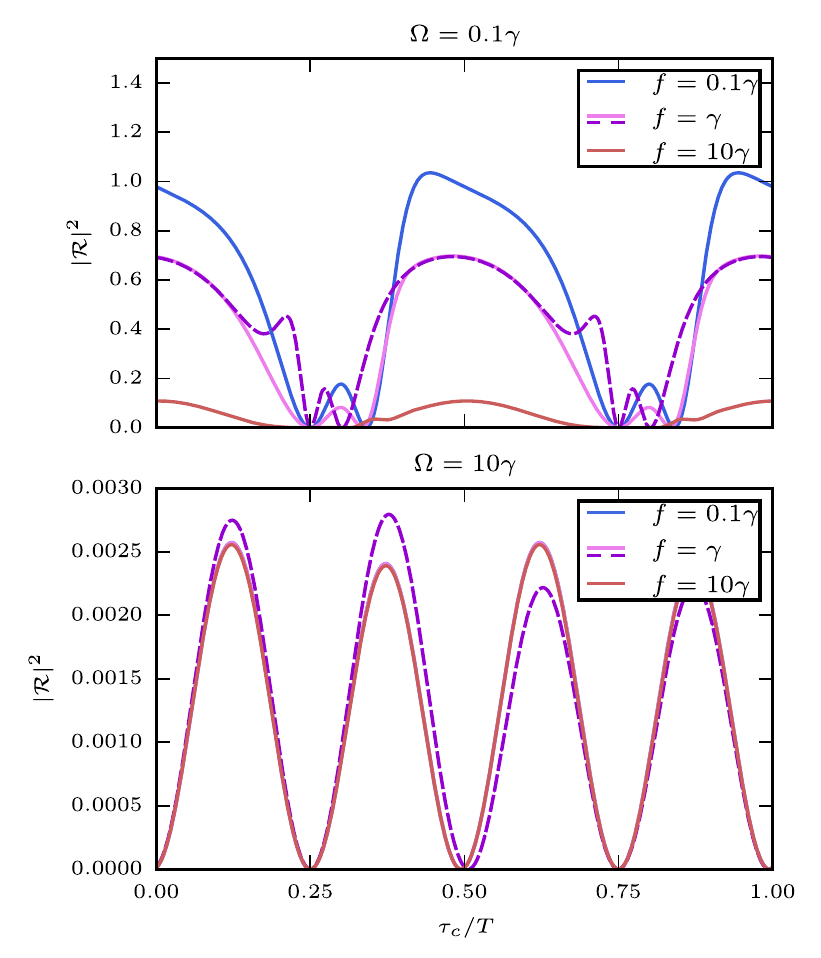}
	\caption{(color online) Reflection from the qubit on resonance, $\delta = 0$, for the time-modulated coupling $g(t) = g_0 \cos \Omega t$ with slow $\Omega =0.1 \gamma$ (top) and fast $\Omega=10 \gamma$ (bottom) modulation frequencies, expressed in the units of $\gamma = \pi g_0^2$. Top: The adiabatic approximation (dashed line) obtained from Eq.~\eqref{eq:adiabatic_sol} deviates from the numerical solution in the vicinities of time instants when the coupling is quenched. For all values of $f$, the reflection is suppressed at these points. At large $f$, the reflection is completely suppressed because of the qubit's saturation. Bottom: The high-frequency modulation suppresses the reflection for any input power $f$. The numerical result (solid lines) is well approximated by the high-frequency result (dashed line) obtained from Eqs.~\eqref{eq:rhoav0}-\eqref{eq:rhoav1}. }
	\label{fig:qubitreflection}
\end{figure}

The numerically obtained reflection $|\mathcal{R}|^2(\tau_c)$ in the quasi-stationary limit with a cosinusoidal modulation of  $g (t)$ is shown in Fig.~\ref{fig:qubitreflection} for different input powers $f$ on a single period $T$. The results for weak input powers are equivalent to those obtained by the Floquet scattering approach in~\cite{Pletyukhov17}. This is confirmed analytically by perturbatively evaluating Eq.~\eqref{ct0} in the weak power limit $f \ll |\bar{\gamma}- i \bar{\delta}|$. Obtaining
\begin{align*}
	s_{qs,2} (t) 
	& \approx 
		-i \sqrt{\pi f} e^{- F (t)} \int_{-\infty}^t d t'  e^{F (t')} g^* (t'),
\end{align*}
with $F (t) = \int_0^{t} d t' [\gamma (t')-i \delta (t')]$, we exactly reproduce the reflection amplitude given in Eq.~(34) of Ref.~\cite{Pletyukhov17}.

We note that for the  modulation protocol $g (t) = g_0 \cos \Omega t$, where the coupling periodically switches its sign, the period of the quasi-stationary reflection (shown in Fig.~\ref{fig:qubitreflection}) is exactly half of the modulation period $T$. Moreover, reflection goes to zero not only at the quench times when $g(t)=0$ but also at some intermediate times. Remarkably, the adiabaticity is violated around the quench points even at sufficiently slow modulation, as one can conclude from the comparison (see the upper panel) of the numerical solution (solid line) for $f=\gamma\equiv \pi g_0^2$ with the corresponding adiabatic approximation of section~\ref{sect:adiab} (dashed line). This feature has already been noticed previously in~\cite{Pletyukhov17} for weak input powers $f$, and now we see that it persists with increasing $f$. In the beginning ($\tau_c \approx 0$), in the middle ($\tau_c \approx T/2$), and in the end ($\tau_c \approx T$) of the modulation period, the instantaneous relaxation rate $\gamma (t)$ is larger than $\Omega$, and the adiabatic approximation approaches the numerical result. The overall decrease of the reflection with increasing $f$ is naturally associated with the qubit's saturation. 

In contrast to the adiabatic approximation, the high-frequency approximation at fast modulations, introduced in section~\ref{sec:high_fr}, is most accurate in the vicinities of the quench points, as follows from its comparison (dashed line) with the numerical solution (solid lines) in the lower panel of Fig.~\ref{fig:qubitreflection}. In general, fast modulation tends to suppress the reflection for any value of the input power $f$. 

\subsection{Power spectrum}

The power spectrum is related to the correlation function of outgoing photons
\begin{subequations}
\label{eq:g1alpha}
\begin{align}
g_\alpha^{(1)}(\tau,\tau_c) &= \langle a_{\alpha,\textrm{out}}^{\dagger}  (\tau_c+\tau) a_{\alpha,\textrm{out}} (\tau_c)  \rangle  \nonumber \\
	&= \phantom{+} \delta_{\alpha,L} e^{i \omega_0 \tau} f \mathcal{R}^* (\tau_c + \tau) \mathcal{R} (\tau_c) \label{corr1elL} \\
	&\phantom{=} + \delta_{\alpha,R} e^{i \omega_0 \tau} f \mathcal{T}^* (\tau_c + \tau) \mathcal{T} (\tau_c) \label{corr1elR} \\
	&\phantom{=} + \phantom{\delta_{\alpha,R}} e^{i \omega_0 \tau}  \pi g^* (\tau_c + \tau ) g (\tau_c)  G_1 (\tau , \tau_c), \label{corr1inel} 
\end{align}
\end{subequations}
where the terms~\eqref{corr1elL} and~\eqref{corr1elR} give rise to the elastic contribution to the power spectrum for reflected and transmitted photons, respectively, while the common term~\eqref{corr1inel} constitutes the inelastic contribution.

\begin{figure}[t!]
	\centering
	\includegraphics[width=\columnwidth]{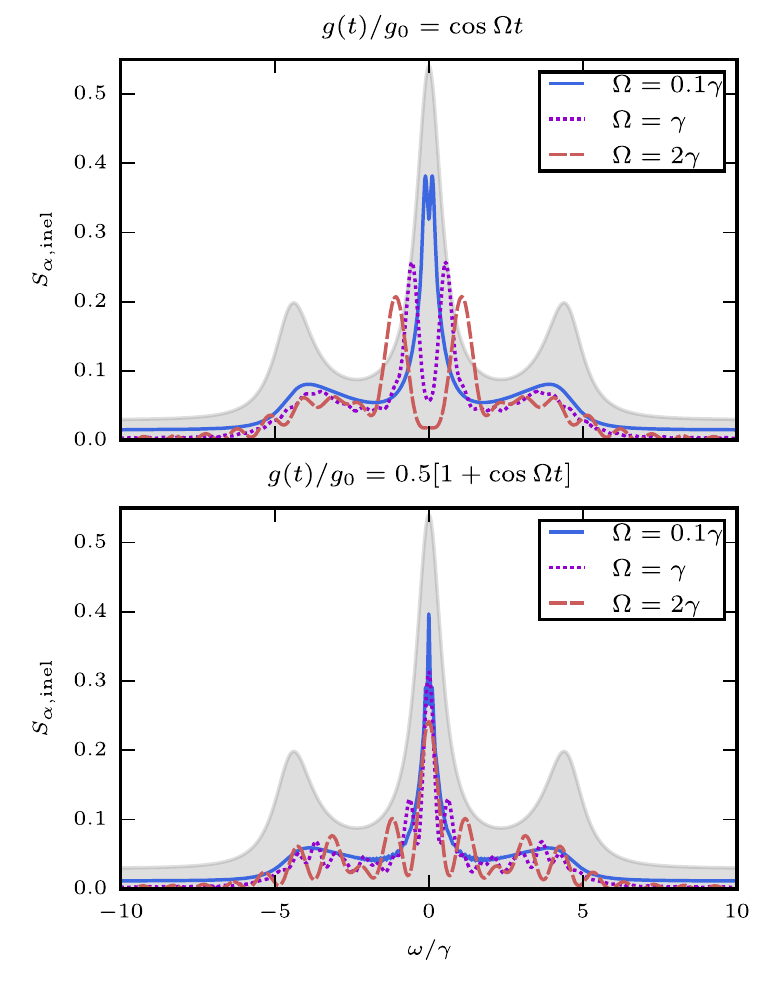}
	\caption{(color online) Inelastic power spectra of the qubit strongly driven $(f = 200\gamma)$ on resonance ($\delta = 0$) for the (top) sign-change protocol $g = g_0 \cos \Omega t$ and the (bottom) on-off protocol $g (t) = g_0 (1+ \cos \Omega t)$ for various modulation frequencies $\Omega$. The Mollow triplet for the corresponding stationary case at coupling $g = g_0$ is shown for comparison in grey. Additional broadened peaks consistent with the Floquet spectrum arise due to the periodic modulation and may destructively interfere as is for example seen in the missing main peak in the case of the sign-change protocol (top). Similar features in the power spectrum have been reported in~\cite{Gustin17} for a qubit subject to a pulsed excitation.}
	\label{fig:qubit_mollow}
\end{figure}
A proper definition of the power spectrum requires time-translational invariance, which can be restored in the periodic quasi-stationary limit by averaging the variable $\tau_c \in [0,T]$ over a period of modulation. With the Fourier expansion $\mathcal{R}(\tau_c) = \sum_m \mathcal{R}^{(m)} e^{-im \Omega \tau_c}$ of the quasi-stationary reflection (and, equivalently, transmission) amplitude, we hence obtain for the elastic contribution
\begin{align}
S_{L,\textrm{el}} (\omega) &= \frac{1}{2\pi} \int_{-\infty}^\infty d\tau \left[ \frac{1}{T} \int_0^T d\tau_c \, g^{(1)}_{L,\textrm{el}}(\tau,\tau_c) \right] e^{-i\omega\tau} \nonumber \\
	&= f \sum_m |\mathcal{R}^{(m)}|^2 \, \delta(\omega - \omega_0 - m\Omega),
\end{align}
which peaks not only at the working frequency $\omega_0$, but also at frequencies shifted from $\omega_0$ by integer multiples of $\Omega$.  An analogous expression holds for $S_{R,\textrm{el}} (\omega)$ of the transmitted photons with the replacement $\mathcal{R} \to \mathcal{T}$.

Evaluation of the inelastic contribution~\eqref{corr1inel} to the power spectrum requires knowledge of the vector $\vec{G} (\tau , \tau_c)  = \langle \hat{\vec{s}} (\tau_c +\tau) \hat{s}_2 (\tau_c) \rangle - \vec{s}_{\textrm{qs}} (\tau_c +\tau) s_{\textrm{qs},2} (\tau_c)$ in the quasi-stationary limit. We note that only its first component is required, which has the property $G_1(-|\tau|,\tau_c) = G_1(|\tau|,\tau_c)^*$. It is thus sufficient to find an equation for $\vec{G}(\tau,\tau_c)$ by means of the quantum regression theorem for $\tau > 0$ only. It reads
\begin{align}
	\frac{d}{d \tau} \vec{G} (\tau , \tau_c) &= A (\tau_c+\tau) \vec{G} (\tau , \tau_c) ,
	\label{eqG}
\end{align}
and its solution can be expressed in terms of $O(t)$ which is governed by the same periodic matrix $A(t)$,
\begin{align}
	\vec{G} (\tau , \tau_c) &= O (\tau +\tau_c) O^{-1} (\tau_c) \vec{G}^{(0)} (\tau_c). \label{GSol}
\end{align}
For the initial condition $\vec{G} (0,\tau_c) = \vec{G}^{(0)} (\tau_c)$, we employ the quasi-stationary values of $\vec{s}_{\textrm{qs}} (\tau_c)$, setting
\begin{align}
\vec{G}^{(0)}(\tau_c)   &= \begin{pmatrix}
							\frac12 s_{\textrm{qs},3} (\tau_c) - |s_{\textrm{qs},2} (\tau_c)|^2 \\
							- s_{\textrm{qs},2}^2 (\tau_c) \\
							- s_{\textrm{qs},3} (\tau_c) s_{\textrm{qs},2} (\tau_c)
						  \end{pmatrix}. \label{Ginit}
\end{align}
From the representation $O(t)=P(t) e^{Bt}$ we find for the inelastic contribution
\begin{align}
g^{(1)}_{\alpha,\textrm{inel}}(\tau,\tau_c) &= \phantom{+} \pi e^{i\omega_0 \tau} \Theta(\tau) \vec{V}_+(\tau_c + \tau) \cdot e^{B\tau} \vec{V}_0(\tau_c) \nonumber \\
	&\phantom{=} + \pi e^{i\omega_0 \tau} \Theta(-\tau) \left[ \vec{V}_+(\tau_c) \cdot e^{-B\tau} \vec{V}_0(\tau_c + \tau) \right]^*
\end{align}
with the periodic vector functions
\begin{align}
\vec{V}_+(\tau_c) &= g^*(\tau_c) P(\tau_c) \vec{n}_1, \\
\vec{V}_0(\tau_c) &= g(\tau_c) P^{-1}(\tau_c) \vec{G}^{(0)}(\tau_c),
\end{align}
and $\vec{n}_1 = (1,0,0)^T$.

As before, we insert the Fourier expansions for the periodic vectors $\vec{V}_{+,0}(\tau_c) = \sum_m \vec{V}_{+,0}^{(m)} e^{-im\Omega \tau_c}$ to evaluate the $\tau_c$-average over a single period of modulation to restore time-translational invariance. Additionally, it is useful to express the matrix $B = \sum_{j=1}^3 b_j \, \vec{\chi}_r^{(j)}\otimes \vec{\chi}_l^{(j)}$ in terms of its eigenvalues $b_j$ and the corresponding biorthonormal left and right eigenvectors obeying $\vec{\chi}_l^{(j)} \cdot \vec{\chi}_r^{(j')} = \delta_{jj'}$. This gives direct analytical access to resonance positions $\omega_{m,j} = \omega_0 + m\Omega + \mathrm{Im} \, b_j $ and widths $\sigma_j = -\mathrm{Re} \, b_j$ in the inelastic power spectrum 
\begin{align}
& S_{\alpha,\textrm{inel}} (\omega) \nonumber \\
&\quad = \frac{1}{2\pi} \int_{-\infty}^\infty d\tau \left[ \frac{1}{T} \int_0^T d\tau_c \, g^{(1)}_{\alpha,\textrm{inel}}(\tau,\tau_c) \right] e^{-i\omega\tau} \nonumber \\
	&\quad =  \sum_m \sum_{j=1}^3 \mathrm{Re} \left[\frac{(\vec{V}_+^{(-m)} \cdot \vec{\chi}_r^{(j)}) (\vec{\chi}_l^{(j)} \cdot \vec{V}_0^{(m)})}{i (\omega - \omega_0 - m \Omega - \mathrm{Im} \, b_j) - \mathrm{Re} \, b_j}\right].
\end{align}
This result indicates equidistant additional resonances introduced by the periodic modulation which can be understood from a dressed state picture: The periodic modulation further splits the dressed states of the qubit driven through the transmission line into $m$ equidistant Floquet modes. Numerical results shown in Fig.~\ref{fig:qubit_mollow} confirm this behavior but also show that for the modulation protocol $g(t) = g_0 \cos \Omega t$ some of the resonances are suppressed and the main peak splits into two side-peaks. This behavior can in principle be used for frequency shifting and engineering correlated states of light.

As a final consistency check, let us confirm the power conservation, i.e. that the output photon fluxes $f_\alpha(\tau_c) = g^{(1)}_{\alpha}(0,\tau_c)$ average over one period of modulation to give the input flux $f = \bar{f}_L + \bar{f}_R$. In the formal expression, we need to prove the identity
\begin{align}
f &\overset{!}{=} \frac{1}{T} \int_0^T d\tau_c \left[ f_L (\tau_c) + f_R (\tau_c) \right] \nonumber \\
  &= f - \frac{1}{2T} \int_0^T d\tau_c \, \left[ \sum_{j=1}^3 A_{3,j}(\tau_c) s_{\textrm{qs},j}(\tau_c) + C_{3}(\tau_c) \right] \nonumber \\
  &= f - \frac{1}{2T} \int_0^T d\tau_c \, \dot{s}_{\textrm{qs},3}(\tau_c), 
\end{align}
which is indeed fulfilled due to the periodicity of $s_{\textrm{qs},3} (\tau_c)$.

\subsection{Second-order coherence function}

Statistical properties of scattered photons can be analyzed with help of the second-order coherence function
\begin{align}
&  g_{\alpha \alpha}^{(2)} (\tau , \tau_c) = \frac{\langle a_{\alpha,\textrm{out}}^{\dagger} (\tau_c) \, n_{\alpha,\textrm{out}} (\tau_c+\tau) \, a_{\alpha,\textrm{out}} (\tau_c) \rangle}{ f_{\alpha} (\tau + \tau_c) f_{\alpha} (\tau_c)},
\label{eq:g2}
\end{align}
where $n_{\alpha,\textrm{out}} = a_{\alpha,\textrm{out}}^\dagger a_{\alpha,\textrm{out}}$ is the outgoing photon number in channel $\alpha$. This function has been studied earlier in the context of Floquet scattering theory~\cite{Pletyukhov17} for weak input powers $f$, and here we extend those results to larger values of $f$, for which the scattering theory becomes impractical. 

\begin{figure}[t]
	\centering
	\includegraphics[width=0.9\columnwidth]{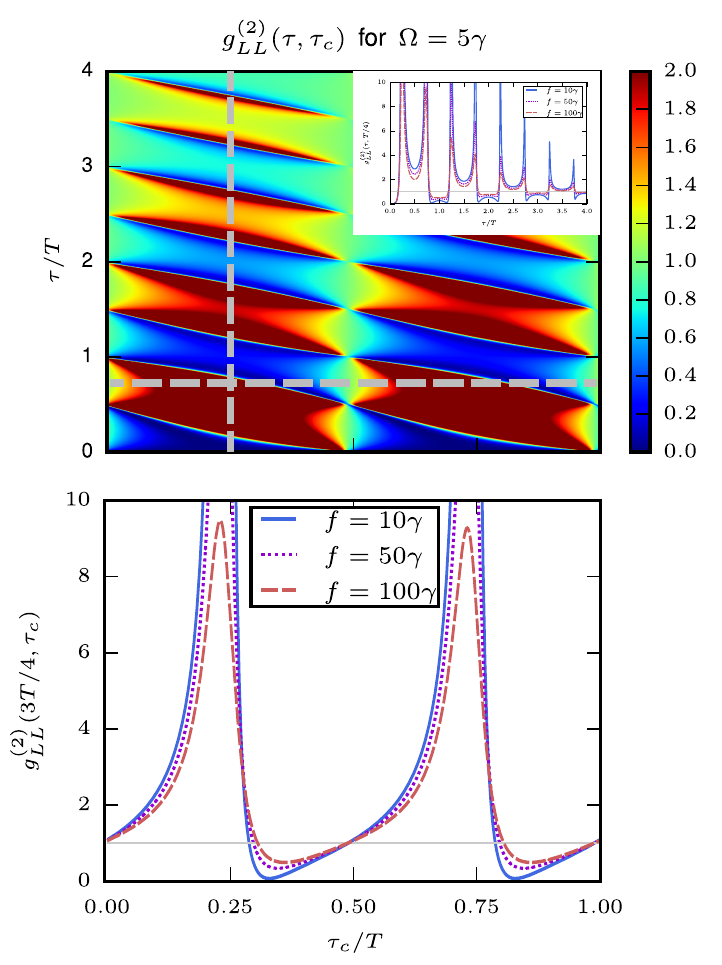}
	\caption{(color online) Upper panel: Second order coherence function $g^{(2)}_{LL}(\tau, \tau_c)$ for the sign-change protocol $g(t)=g_0 \cos (\Omega t)$ of the moderately driven $(f=10\gamma)$ qubit on resonance $(\delta=0)$ at fast modulation frequency $\Omega = 5\gamma$. The oscillations decay with the delay time $\tau$ at the rate $\gamma$ as can be seen e.g. along the vertical cut at $\tau_c = T/4$ shown in the inset. Thus, this rapidly changing behavior takes place only  for sufficiently fast modulations. Lower panel: The strong oscillations in time $\tau_c$ between bunching and anti-bunching behavior reported in~\cite{Pletyukhov17} become less pronounced with increasing input power $f$ as the qubit becomes saturated. The delay time $\tau$ is fixed at the value $3T/4$, which corresponds to the horizontal cut in the top figure. }
	\label{fig:qubitg2}
\end{figure}

Similar to the procedure of evaluating the power spectrum, the quantum regression theorem allows us to write the functions~\eqref{eq:g2} in terms of the vector
\begin{align}
\vec{J} (\tau , \tau_c) &= \langle \hat{s}_1 (\tau_c) \hat{\vec{s}} (\tau+\tau_c) \hat{s}_2 (\tau_c) \rangle \nonumber \\
	&\phantom{=} \qquad - \frac12 s_{\textrm{qs},3} (\tau_c)  \vec{s}_{\textrm{qs}} (\tau +\tau_c),
\end{align}
obeying the same differential equation~\eqref{eqG} in the variable $\tau$ as $\vec{G}(\tau,\tau_c)$ obeys, but with the initial conditions $\vec{J}^{(0)} (\tau_c)  = -\frac12 s_{\textrm{qs},3} (\tau_c) \vec{s}_{\textrm{qs}} (\tau_c)$. We find
\begin{subequations}
\begin{align}
g_{LL}^{(2)} (\tau , \tau_c) &= 1 + \frac{2 \mathrm{Re} [\nu_3 (\tau_c) \nu_3 (\tau +\tau_c) J_3 (\tau, \tau_c)]}{f_L (\tau +\tau_c ) f_L (\tau_c) }  \\
g_{RR}^{(2)} (\tau , \tau_c) &= 1 + \frac{2 \mathrm{Re} [ \nu_3 (\tau_c)  \vec{\nu} (\tau +\tau_c) \cdot \vec{J} (\tau , \tau_c) ]}{ f_R (\tau+\tau_c) f_R (\tau_c)} \nonumber \\ 
&\phantom{= 1 \,\,} + \frac{ 2 \mathrm{Re} [ \nu_2(\tau_c)  \vec{\nu} (\tau +\tau_c) \cdot \vec{G} (\tau, \tau_c)] }{ f_R (\tau+\tau_c) f_R (\tau_c)},
\end{align}
\end{subequations}
where $\vec{\nu}(t) = (0,0,\frac14 \gamma(t))^T - \vec{C}^*(t)$ is a modification of the vector~\eqref{Ct}.

As shown in Fig.~\ref{fig:qubitg2}, the oscillations between strong bunching and anti-bunching behavior observed in~\cite{Pletyukhov17} become less pronounced as the input power $f$ is increased, see the bottom panel of this figure, corresponding to the horizontal (dashed grey) cut  in the upper panel. This behavior can again be attributed to the qubit's saturation. For fast enough modulation frequencies ($\Omega \gtrsim \gamma$), the bunching peaks remain sizeable on the range of several $T$ in the delay time $\tau$ even for the moderate input power $f = 10\gamma$ , see the inset of the upper panel corresponding to the vertical (dashed grey) cut of the contour plot.

Thus, the rapid bunching-to-antibunching changes in behavior of the $g^{(2)}$ function, which result from the system's non-adiabatic response to an external modulation and which have been predicted in~\cite{Pletyukhov17} for weak input powers $f$, appear to persist in a broad range of input power $f$. We observe that the positions of the bunching peaks remain insensitive to $f$, and only their heights gradually go down with increasing $f$.

\begin{figure}[b]
	\centering
	\includegraphics[width=\columnwidth]{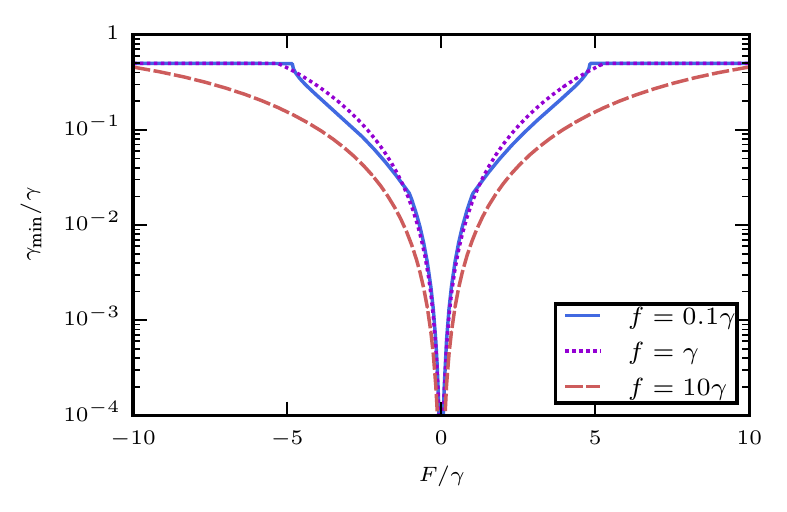}
	\caption{(color online) Smallest dissipation rate $\gamma_{\textrm{min}}$ of the $\Lambda$-system as a function of the classical drive field amplitude $F$. Instead of directly modulating the coupling strength $g$, the three-level $\Lambda$-system allows tuning of $\gamma_{\textrm{min}}$ by means of $F$. The input power $f$ of the probe field has little effect on this behavior.}
	\label{fig:eit_spec}
\end{figure}

\section{Driven $\Lambda$-system}\label{sec:eit}

Next, we consider a three-level system in the $\Lambda$-scheme where direct transition from the ground state $|g\rangle$ to an intermediate metastable state $|s\rangle$ is forbidden. Such systems are known to exhibit electromagnetically induced transparency (EIT), an effect which has first been observed in atomic vapors~\cite{KochEIT86,BollerEIT91}. Recently, this phenomenon has also been demonstrated in superconducting circuits~\cite{AbduEIT10} thus paving the way for potential applications in quantum information processing.

The drive field at frequency $\omega_d$, which is  nearly resonant with frequency $(\omega_e - \omega_s)$ of the transition $|s\rangle \to |e\rangle$ to the excited state, is conventionally treated classically. Our interest lies in a time-modulation of the drive amplitude $F(t)$ causing a periodic switching between opaque and transparent behavior of this system upon irradiation of the coherent probe field $|\Psi_0 \rangle$ at frequency $\omega_0$, which is nearly resonant with frequency $\omega_e$ of the transition $|g\rangle \to |e\rangle$. This model is described by the Hamiltonian
\begin{align}
H(t) &= \omega_e |e \rangle \langle e | + \omega_s |s \rangle \langle s| + \left[ F(t) e^{-i \omega_d t} |e \rangle \langle s | + h.c. \right] \nonumber \\ 
  &\phantom{=} \quad + H_{w} + \sum_\alpha \int d\omega \left[\frac{g}{\sqrt{2}} a^\dagger_{\alpha\omega} | g \rangle \langle e | + h.c.\right].
\end{align}
In the following, we show that this system exhibits non-adibatic effects similar to those of the two-level system with a modulated coupling strength. At the same time, the $\Lambda$-scheme with a periodically modulated drive field  is more feasible for an experimental realization.

Dissipative dynamics of the $\Lambda$-system in the co-rotating frame is governed by the master equation \eqref{eq:dot_rhop} with the
matrix
\begin{widetext} 
\begin{align}
A(t) = 
\begin{pmatrix}
- \gamma & 0 & - i  \sqrt{\gamma f/2} &  i  \sqrt{\gamma f/2} & - i F(t) & i F^*(t) & 0 & 0 \\
0 & 0 & 0 &  0 &  i F(t) & - i F^*(t) & 0 & 0 \\
-2 i  \sqrt{\gamma f/2} & - i  \sqrt{\gamma f/2} & - i (\delta_1 - i \gamma/2) & 0 & 0 &0 & i F^*(t) & 0 \\
2 i \sqrt{\gamma f/2} & i  \sqrt{\gamma f/2} & 0 & i (\delta_1 +i \gamma/2) & 0 & 0 & 0 & - i F(t) \\
- i F^*(t) & i F^*(t) & 0 &0 & - i (\delta_2 - i \gamma/2) & 0 & 0 & i \sqrt{\gamma f/2}  \\
i F(t) & - i F(t) & 0 & 0 & 0 & i (\delta_2 + i \gamma/2) & -i \sqrt{\gamma f/2}  & 0 \\
0 & 0 & i F(t) & 0 & 0 & - i  \sqrt{\gamma f/2} & - i (\delta_1 - \delta_2) & 0 \\
0 & 0 & 0 & -i F^*(t) & i  \sqrt{\gamma f/2} & 0 & 0 & i (\delta_1 - \delta_2)
\end{pmatrix},
\end{align}
and the vector 
\begin{align}
\vec{C} = (0,0,i \sqrt{\gamma f/2}, -i \sqrt{\gamma f/2} ,0,0,0,0)^T ,
\end{align}
which are written in the basis
\begin{align}
\vec{s} (t) = \langle \hat{\vec{s}}(t) \rangle = ( \langle P_e(t) \rangle, \langle P_s(t) \rangle, \langle \tilde{\sigma}_+^{(g)} (t)\rangle, \langle \tilde{\sigma}_-^{(g)} (t)\rangle,  \langle \tilde{\sigma}_+^{(s)} (t) \rangle, \langle \tilde{\sigma}_-^{(s)} (t)\rangle,  \langle \tilde{\sigma}_+^{(r)} (t) \rangle, \langle \tilde{\sigma}_-^{(r)} (t)\rangle)^T. \nonumber
\end{align}
\end{widetext}
Here, $P_e = |e\rangle \langle e|$, $P_s = |s\rangle \langle s |$, $\sigma_-^{(g)} =| g \rangle \langle e |$, $\sigma_-^{(s)} =| s \rangle \langle e |$, $\sigma_-^{(r)} =| g \rangle \langle s |$, $\sigma_+^{(g,s,r)} = ( \sigma_-^{(g,s,r)} )^\dagger$ and the tildes indicate expectation values to be evaluated in the co-rotating frame analogous to the two-level system.  Additionally, we have defined the detunings $\delta_1 =\omega_0 - \omega_e$ and $\delta_2 =\omega_d - (\omega_e - \omega_s)$, and the bare dissipation rate $\gamma = \pi |g|^2$. Note that for a computation of the transmission amplitude $\mathcal{T}$ one can use~\eqref{Rt} and~\eqref{Tt} with $\langle \tilde{\sigma}_- \rangle \to \langle  \tilde{\sigma}_-^{(g)} \rangle $.

\begin{figure}[b]
	\centering
	\includegraphics[width=\columnwidth]{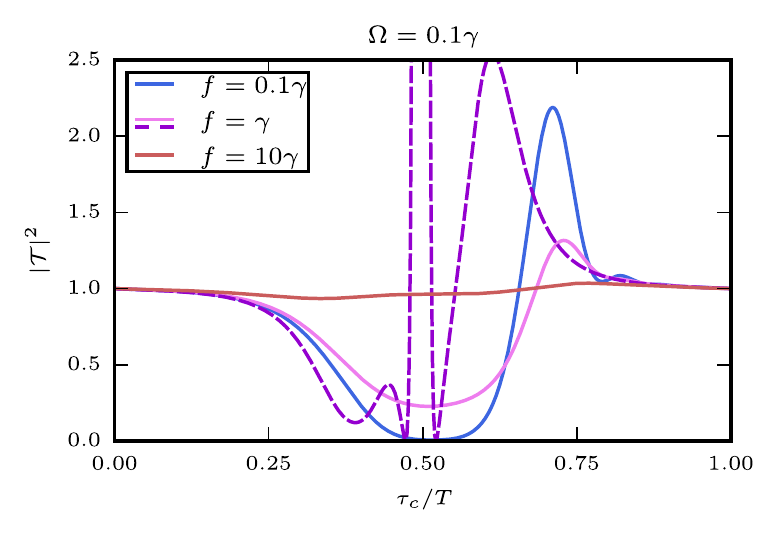}
	\caption{(color online) Transmission through the $\Lambda$-system which is driven on resonance $(\delta_1 = \delta_2 = 0)$ by both probe and drive pulses. The drive field has a periodically modulated amplitude $F(t) = 10 \gamma \, [1 + \cos \Omega t]$  at $\Omega = 0.1 \gamma$. The adiabatic approximation (dashed line) is only valid far away from the critical region defined by $\Omega > \gamma_{\textrm{min}}$ (cf. Fig.~\ref{fig:eit_spec}). In the time window where it breaks down, the system responds non-adiabatically. At large powers $f$ of the probe field, these effects are, however, washed out because of the system's saturation.}
	\label{fig:eit_refl}
\end{figure}

Unlike in the two-level system, dissipation rates of the $\Lambda$-system depend on multiple parameters. At fixed $\gamma$, the smallest dissipation rate $\gamma_{\textrm{min}}$ has a nearly quadratic parametric dependence on the drive amplitude $F$, as shown in Fig.~\ref{fig:eit_spec}. This indicates that we can push the system into the non-adiabatic regime with $\Omega > \gamma_{\textrm{min}}$ by sweeping the values of $F$ towards zero. Note that $\gamma_{\textrm{min}}$ shows little sensitivity to the intensity $f$ of the probe field.

In the EIT model with constant $F \neq 0$, the system is fully transparent on resonance $\delta_1 = \delta_2 =0$ leading to $|\mathcal{T}|^2=1$. When $F$ is momentarily quenched, the metastable state $| s \rangle $ is decoupled for a short while, and the remaining two-level system $\{ | g \rangle , | e \rangle \}$ tends to develop full reflection (and, hence, zero transmission), provided that the probe field does not saturate the system. In the next time-instant, the state $|s \rangle$ is re-coupled again, which leads to non-adiabatic changes in transmission properties. Changing $F$ periodically in time, e.g. by $F (t) =10 \gamma [1+ \cos \Omega t ]$ can thus result in a quasi-stationary behavior of the transmission with large deviations from unity on a single period of modulation. This  is illustrated in Fig.~\ref{fig:eit_refl}. Switching between opaqueness and transparency closely resembles the behavior of the two-level system where the modulated coupling effectively performs the function similar to that of $F$, though with the reciprocal effect. As it is seen from the comparison of  the adiabatic approximation (dashed line) with the numerical solution (solid line) at $f = \gamma$, the system's response is non-adiabatic during a large part of the period for rather slow modulation frequency $\Omega =0.1 \gamma$. This behavior is due to the modulation protocol of $F$ which deeply penetrates into the critical region defined by $\Omega > \gamma_{\textrm{min}}$ (cf. Fig.~\ref{fig:eit_spec}).

In the high-frequency regime of modulation, the regular EIT effect with unit transmission on resonance is again restored as long as the time average $\bar{F} \neq 0$. For $\bar{F} =0$ we obtain an effective decoupling of the metastable state $| s \rangle$, reproducing the transmission of the unmodulated two-level system. These conclusions are also supported by the high-frequency expansion~\eqref{eq:high_fr}.

\section{Driven Kerr non-linearity system}\label{sec:kerr}

In the third  application of our formalism, we consider the driven Kerr nonlinearity model. It consists of a single cavity mode $b$ with an effective local photon-photon interaction $U$, which is coupled to the transmission line. Its dissipative dynamics in the co-rotating frame is governed by the Lindblad master equation~\eqref{eq:lind} with $O =b$ and 
\begin{align}
H_{\textup{eff}} (t) &= - \delta (t)  b^\dagger b + \frac{U}{2} b^\dagger b^\dagger b b + \sqrt{\gamma f} (b+b^{\dagger}).
\end{align}
In the following, we consider time-modulation of the detuning $\delta= \omega_0 - \omega_e$, where $\omega_e$ is the cavity frequency.

\begin{figure}[b!]
	\centering
	\includegraphics[width=\columnwidth]{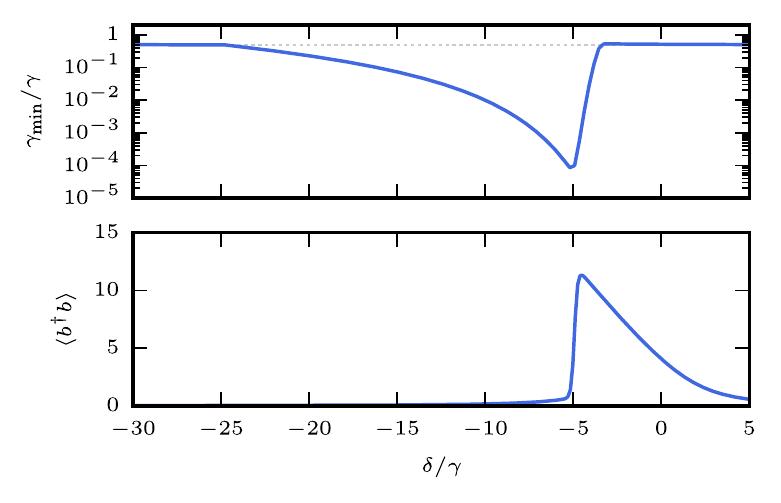}
	\caption{(color online) Upper panel: Smallest dissipation rate $\gamma_{\textrm{min}}$ of the Kerr nonlinearity model as a function of detuning $\delta$ for $U=-\gamma/2$ and $f = 16\gamma$. Similar to the two- and three-level systems, the smallest dissipation rate is significantly suppressed within the critical region, though it remains finite. Lower panel: One of the important signatures of the dissipative phase transition is a prominent increase in the stationary occupation number $\langle b^\dagger b \rangle$ which peaks at the same parameter value for which the minimal dissipation rate is reached. In the following, periodic modulation of $\delta$ is considered across the whole critical region with modulation frequency $\Omega \ll \gamma$.}
	\label{fig:kerr_steady}
\end{figure}

Before turning to the time-dependent case, let us revisit the steady state results obtained by Drummond and Walls~\cite{Drummond80} and recently extended to include two-photon driving~\cite{BartoloCiuti}. The dissipative phase transition that this system exhibits for large $f \gg \gamma$ and small $|U| \ll \gamma$ has numerous manifestations. Experimentally, the most feasible quantity is the steady state occupation $\langle b^\dagger b \rangle$. Sweeping  detuning $\delta $ over the  bistability critical region (i.e., where the corresponding semiclassical solution has multiple solutions), one can observe a strong  enhancement in the occupation number (shown in Fig.~\ref{fig:kerr_steady}, bottom). Away from this region, $\langle b^\dagger b \rangle$ decays to small values. The peak value rapidly grows with increasing ratio $f/U^2$.  This behavior goes hand in hand with the entropy of the cavity: Outside of the critical region, the state is a pure coherent state corresponding to zero entropy, but becomes a complicated mixed state within the critical region.

This critical behavior can again be attributed to the smallest dissipation rate $\gamma_{\textrm{min}}$ being significantly suppressed (shown in Fig.~\ref{fig:kerr_steady}, top), a phenomenon which is also known as the critical slowing down. In fact, the Liouvillian gap does not completely close. The minimal value of the dissipation is reached at $\delta$ where also the occupation number peaks.

Of particular interest is a periodic modulation of parameters which drives the system in and out of the critical region. Recently, it has been proposed~\cite{Casteels16} that in this way one can dynamically simulate a hysteretic behavior in the Kerr model, which has been experimentally observed~\cite{Rodriguez17} in the corresponding setup soon after. Interestingly, the hysteresis-like behavior follows the stable branches of the semiclassical mean-field solution rather than the exact steady state quantum solution. An explanation of this property has been provided in the context of the driven-dissipative Rabi model~\cite{Boite17} where it has been shown that long-lived metastable states with a small effective decay rate prevent reaching the true steady state. As pointed out in Ref.~\cite{Casteels16}, this goes together with a breakdown of adiabaticity, which we have also seen in the previously discussed models. The studies cited above give strong indications that such behavior seems to be common for all systems featuring dissipative phase transitions.

\begin{figure*}[t!]
	\centering
	\includegraphics[width=\textwidth]{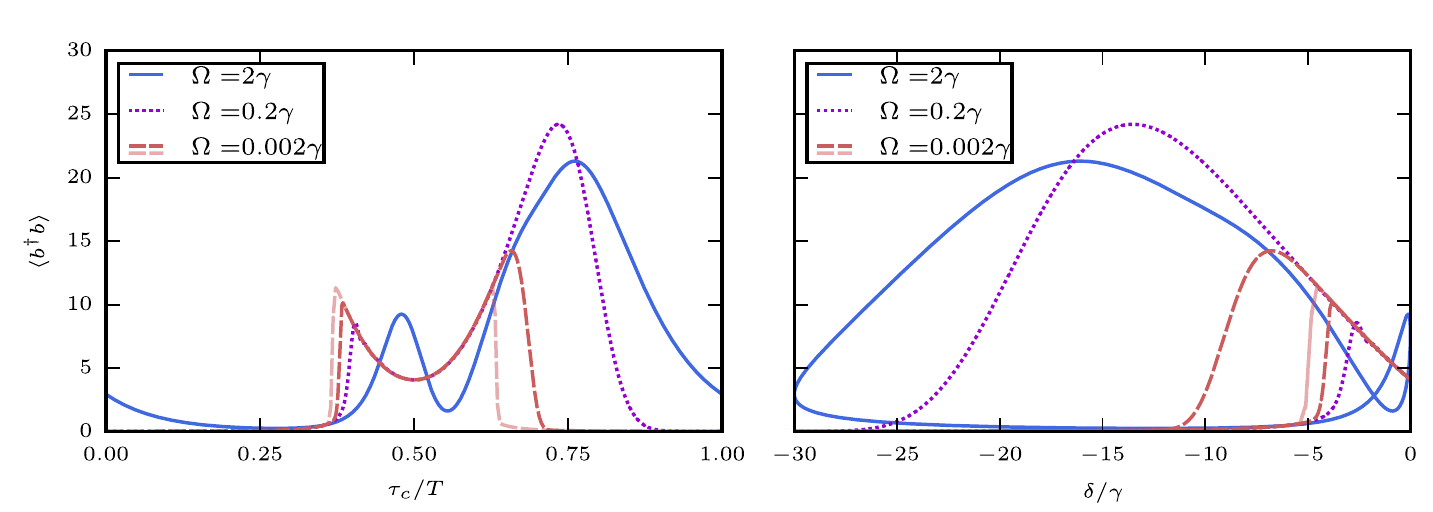}
	\caption{(color online) Left: quasi-stationary occupation $\langle b^{\dagger} b \rangle (t)$ of the Kerr nonlinearity model under periodic modulation of detuning $\delta(t) = -15 \gamma[1+\cos \Omega t]$ for various modulation frequencies $\Omega$. Right: the same dependence in the parametric representation. For moderate modulation frequencies, the occupation is strongly enhanced compared to the true steady state shown in the lower panel of Fig.~\ref{fig:kerr_steady}. The adiabatic approximation based on Eq.~\eqref{eq:adiabatic_sol} is given by the light brown curve (note that in the parametric representation it lies very close to the stationary state result). The numerical results for rather small  frequency $\Omega = 0.002 \gamma$ (brown) still drastically deviate from the corresponding adiabatic approximation (light brown). As discussed by Casteels \textit{et al.}~\cite{Casteels16}, the dynamical hysteresis seen in the parametric plot is directly related to the breakdown of adiabaticity in the critical region, and the hysteresis area depends on the width of the parameter range where $\Omega > \gamma_{\textrm{min}}$.}
	\label{fig:kerr_td}
\end{figure*}

In contrast to modulating $f$ discussed in Ref.~\cite{Casteels16}, we choose to vary in time the parameter $\delta$. This is advantageous since one can sweep {\it in} and {\it out} of the critical region in positive and negative sweep direction, starting on both sides from non-critical regions characterized by zero values of entropy. In particular, we have found that it is hard to ensure this when sweeping $f$ at fixed $\delta$. Our modulation protocol is designed to cover the whole critical region, namely $\delta(t)= -15\gamma[1+\cos\Omega t]$ for the parameters $U=-\gamma/2$ and $f = 16 \gamma$. 

The quasi-stationary occupation over a single period and its parametric dependence on the parameter $\delta$ is shown in Fig.~\ref{fig:kerr_td}. The left panel reveals a clear rise in occupation whenever $\delta$ is deep inside the critical region, which is followed by an exponential drop. Note that in comparison with the steady state result, the occupation is significantly enhanced for intermediate modulation frequency $\Omega = 0.2 \gamma$ (dotted line).  Further increase of $\Omega$ up to the value $2 \gamma$ does not enhance the occupancy any more (solid line); moreover, hysteretic properties are not seen any more in the parametric representation of the right panel.

For slow modulation, the quasi-stationary solution does not converge to the adiabatic approximation based on Eq.~\eqref{eq:adiabatic_sol} (light brown curve) even for $\Omega = 0.002\gamma$. This points towards the non-adiabatic system's response when its parameters are driven across the region of bistability.

\section{Summary}
Based on Floquet's theorem, we have proposed a representation for the quasi-stationary density operator of a periodically driven-dissipative open quantum system. We have established both adiabatic and high-frequency expansions in a systematic way. Importantly, the corresponding approximations can be efficiently benchmarked against numerical results which are achieved by integration over a single period of modulation. A breakdown of the adiabatic approximation signals the non-adiabatic system's response when it enters the regime of critical slowing down.

We applied the developed formalism to three different models with periodically time-dependent parameters, which all exhibit a temporary suppression of the smallest dissipation rate.

For the two-level system, a modulation of the coupling strength to the transmission line causes significant changes in transmission properties, power spectra and statistical properties of scattered photons. 

For the three-level $\Lambda$-system, a modulation of the classical driving of the metastable state can lead to considerable modifications of the EIT phenomenon.  

In the driven Kerr nonlinearity model, we have studied periodic sweeping of the detuning $\delta$ across the parameter region featuring the driven-dissipative phase transition. We have found that even for slow modulation frequencies non-adiabatic effects dominate, indicating that adiabatic expansions will generally fail in critical parameter regimes of such systems. 

\begin{acknowledgments}
We gratefully acknowledge useful discussions with D. Krimer and M. R. Wegewijs. V. R. is supported by the Deutsche Forschungsgemeinschaft (DFG) under grant RTG 1995. Work of V.G. is part of the Delta-ITP consortium, a program of the Netherlands Organization for Scientific Research (NWO) that is funded by the Dutch Ministry of Education, Culture and Science (OCW). 

\end{acknowledgments}

\end{document}